\documentclass{llncs}
\usepackage{amsmath}
\usepackage{graphicx}

\usepackage{cite}
\RequirePackage{doi}
\usepackage{hyperref}
\usepackage{paralist}
\usepackage[strings]{underscore}

\usepackage[xindy]{glossaries}

\newacronym{ltl}{LTL}{Linear-time Temporal Logic}

\newacronym{ctl}{CTL}{Computation Tree Logic}

\newacronym{ptl}{PTL}{Probabilistic Temporal Logic}

\newacronym{pctl}{PCTL}{Probabilistic Computation Tree Logic}

\newacronym{pctl*}{PCTL*}{Probabilistic Computation Tree Logic*}

\newacronym{tctl}{TCTL}{Timed Computation Tree Logic}

\newacronym{stl}{STL}{Signal Temporal Logic}

\newacronym{ltl-x}{LTL-X}{a version of Linear Time Logic without the `next' operator}

\newacronym{mtl}{MTL}{Metric Temporal Logic}

\newacronym{iactlk-x}{IACTLK-X}{a fragment of the temporal-epistemic logic CTLK, without the `next' operator}

\newacronym{fotl}{FOTL}{First-Order Temporal Logic}

\newacronym{bdi}{BDI}{Belief-Desire-Intention}

\newacronym{prs}{PRS}{Procedural Reasoning System}

\newacronym{are}{ARE}{Autonomous Reasoning Engine}

\newacronym{dmars}{dMARS}{distributed Multi-Agent Reasoning System}

\newacronym{fsm}{FSM}{Finite-State Machine}

\newacronym{pfsm}{PFSM}{Probabilistic Finite-State Machine}

\newacronym{apfsm}{APFSM}{\textit{Autonomous} Probabilistic Finite-State Machine}

\newacronym{pha}{PHA}{Probabilistic Hybrid Automata}

\newacronym{fsa}{FSA}{Finite-State Automata}

\newacronym{ta}{TA}{Timed Automata}

\newacronym{gspn}{GSPN}{Generalised Stochastic Petri Net}

\newacronym{mopn}{MOPN}{Marked Ordinary Petri Net}

\newacronym{ba}{BA}{B\"{u}chi Automata}

\newacronym{mdp}{MDP}{Markov Decision Process}

\newacronym{dtmc}{DTMC}{Discrete-Time Markov Chain}

\newacronym{pars}{PARS}{Process Algebra for Robot Schemas}

\newacronym{fsp}{FSP}{Finite State Processes}

\newacronym{piadl}{$\pi$ADL}{$\pi$ calculus combined with ADL}

\newacronym{csp}{CSP}{Communicating Sequential Processes}

\newacronym{ccs}{CCS}{Calculus of Communicating Systems}

\newacronym{wsccs}{WSCCS}{Weighted Synchronous CCS}

\newacronym{csp|b}{CSP$\|$B}{an integrated formalism combining CSP and B}

\newacronym{fdr}{FDR}{the Failures-Divergences Refinement checker}

\newacronym{jpf}{JPF}{Java PathFinder}

\newacronym{ajpf}{AJPF}{Agent Java PathFinder}

\newacronym{mcmas}{MCMAS}{Model Checker for Multi-Agent Systems}

\newacronym{ltlmop}{LTLMoP}{Linear Temporal Logic MissiOn Planning}

\newacronym{adl}{ADL}{Architecture Description Language}

\newacronym{aadl}{AADL}{Architecture Analysis and Design Language}

\newacronym{lts}{LTS}{Labelled-Transition System}

\newacronym{uml}{UML}{Unified Modelling Language}

\newacronym{sysml}{SySML}{Systems Modelling Language}

\newacronym{robotml}{RobotML}{Robot Modelling Language}

\newacronym{dsl}{DSL}{Domain Specific Language}

\newacronym{sct}{SCT}{Supervisory Control Theory}

\newacronym{ide}{IDE}{Integrated Development Environment}

\newacronym{mde}{MDE}{Model-Driven Engineering}

\newacronym{ai}{AI}{Artificial Intelligence}

\newacronym{fdir}{FDIR}{Fault Detection, Isolation and Recovery}

\newacronym{ifm}{iFM}{Integrated Formal Methods}

\newacronym{ros}{ROS}{Robot Operating System}

\newacronym{dl}{\text{\upshape\textsf{d{\kern-0.05em}L}}}{differential dynamic logic}

\newacronym{vm}{VM}{Virtual Machine}

\newacronym{bip}{BIP}{Behaviour Interaction Priority}

\newacronym{fol}{FOL}{First-Order Logic}

\newacronym{owl}{OWL}{Web Ontology Language}

\newacronym{ria}{RIA}{Restore Invariant Approach}

\newacronym{ocl}{OCL}{Object Constraint Language}

\newacronym{sil}{SIL}{Safety Integrity Level}

\newacronym{rule}{RULE}{Robotic Urban-Like Environment}

\newacronym{forms}{FORMS}{FOrmal Reference Model for Self-adaptation}
\usepackage{wrapfig,lipsum,booktabs}

\begin{document}
\title{A Summary of Formal Specification and Verification of Autonomous Robotic Systems\thanks{Work supported by UKRI Hubs for Robotics and AI in Hazardous Environments: EP/R026092 (FAIR-SPACE), EP/R026173 (ORCA), and EP/R026084 (RAIN). } }

\author{Matt Luckcuck
\and Marie Farrell
\and \\Louise A. Dennis
\and  Clare Dixon
\and Michael Fisher
}
\institute{Department of Computer Science, University of Liverpool, UK \\ \email{ \{m.luckcuck, marie.farrell\}@liverpool.ac.uk }
}

\maketitle 
\begin{abstract}
Autonomous robotic systems are complex, hybrid, and often safety-critical; this makes their formal specification and verification uniquely challenging. Though commonly used, testing and simulation alone are insufficient to ensure the correctness of, or provide sufficient evidence for the certification of, autonomous robotics.  Formal methods for autonomous robotics have received some attention in the literature, but no resource provides a current overview. This short paper summarises the contributions published in~\cite{Luckcuck2019}, which surveys the state-of-the-art in formal specification and verification for autonomous robotics. 
\end{abstract}

\section{Introduction and Methodology}
\label{sec:intro}

This short paper summarises our recently published survey of the formal specification and verification techniques that have been applied to autonomous robotic systems~\cite{Luckcuck2019}, which provides a comprehensive overview and analysis of the state-of-the-art, and identifies promising new research directions and challenges for the formal methods community. Previous work, which draws from this survey, advocates the use of integrated formal methods for autonomous robotic systems~\cite{Farrell2018}.

We define an \textit{autonomous system} as an artificially intelligent entity that makes decisions in response to input, independent of human interaction. \textit{Robotic systems} are physical entities that interact with the physical world. Thus, an \textit{autonomous robotic system} is a machine that uses Artificial Intelligence (AI), has a physical presence in and interacts with the real world. Autonomous robotics are increasingly used in commonplace-scenarios, such as driverless cars~\cite{Fernandes2017}, pilotless aircraft~\cite{webster2011formal}, and domestic assistants~\cite{Dixon2014}. 

For many engineered systems, testing, either by real deployment or via simulation, is deemed sufficient. But autonomous robotics require stronger verification, because of the unique challenges: dependence on sophisticated software control and decision-making, and increasing deployment in safety-critical scenarios. This leads us towards using formal methods to ensure the correctness of, and provide sufficient evidence for the certification of, robotic systems.
 
The corresponding journal paper~\cite{Luckcuck2019} identifies and investigates the following three research questions:
\begin{compactdesc}
\item[RQ1:] What are the challenges when formally specifying and verifying the behaviour of (autonomous) robotic systems?
\item[RQ2:] What are the current formalisms, tools, and approaches used when addressing the answer to \textbf{RQ1}?
\item[RQ3:] What are the current limitations of the answers to \textbf{RQ2} and are there developing solutions aiming to address them?
\end{compactdesc}

To answer these questions we performed a systematic literature survey on \textit{formal modelling of (autonomous) robotic systems}, \textit{formal specification of (autonomous) robotic systems}, and \textit{formal verification of (autonomous) robotic systems}. We restricted our search to papers published from 2007 to 2018, inclusive. 

In addition to answering the research questions, the survey~\cite{Luckcuck2019} illustrates opportunities for research applying formal methods (and \gls{ifm}) to robotics and autonomous systems -- either by identifying the popular languages that integration could use, or by showing the gaps that could be filled by \gls{ifm}. It also provides a brief overview of some popular general software engineering techniques for robotic systems including middleware architectures, testing, and simulation approaches, domain specific languages, graphical notations, and model-driven engineering or XML-based approaches.

\section{Answering the Research Questions}

This section summarises how the results of our survey address the research questions described in \S\ref{sec:intro}. 

To answer \textbf{RQ1}, we identified the challenges describe in the surveyed literature and categorised them as \textit{external} or \textit{internal} to the robotic system. External challenges come from the design and environment, independently of how the system is designed internally. We found two major external challenges in the literature: modelling and reasoning about the system's environment, and providing enough evidence for public trust and regulation. Internal challenges stem from how the system is engineered. The three internal challenges that we found in the literature were related to using: agent-based, multi-robot, and adaptive or reconfigurable systems. These challenges, and the tools and techniques used to overcome them, are discussed at length in \cite[\S 3--4]{Luckcuck2019}.

Tackling internal challenges can have complementary benefits to mitigating external challenges. Reconfigurability is key to safely deploying robots in hazardous environments and much more work needs to materialise in order to ensure the safety of reconfigurable autonomous systems. Therefore, we see a clear link between a robotic system reacting to the changes in its external environment, and reconfigurable systems. Similarly, \textit{rational} agent-based systems that can explain their reasoning provide a good route for providing evidence for public trust or certification bodies. This is because they provide the transparency that is crucial for public trust and certification. 
 A rational agent can provide reasons for its choices, based on input and internal state information.
 \begin{wraptable}{r}{5.3cm}
\vspace{10pt}
\centering
\scalebox{0.85}
{
\begin{tabular}{|c l ||c||c| }
\hline
\multicolumn{2}{|c||}{\textbf{Formalism}} & {\textbf{System}} & {\textbf{Property}} \\
\hline

\multicolumn{2}{|l||}{Set-Based} & 5 &  0 \\ \hline 

\multicolumn{2}{|l||}{State-Transition} & 33 & 0 \\ \hline 

\multicolumn{2}{|l||}{Logics} &  6 & 32 \\ \hline 


	
	
\multicolumn{2}{|l||}{Process Algebra} &  3 &  1 \\ \hline 

\multicolumn{2}{|l||}{Ontology} &  4 & 0 \\ \hline

\multicolumn{2}{|l||}{Other} &  5 & 8 \\ \hline

\end{tabular}}
\caption{\label{tab:formalismSummary} Summary of the types of formalisms for specifying the system and the properties to be checked, summarising~\cite[Table 2]{Luckcuck2019}.}
\vspace{-20pt}
\end{wraptable}

\textbf{RQ2}, asked what are the current formal methods used for tackling the challenges identified by answering \textbf{RQ1}. Thus, we  quantify and describe the formalisms, tools, and approaches used in the literature~\cite[\S 5--6]{Luckcuck2019}, summarised in Table~\ref{tab:formalismSummary} (right). We found that state-transition systems and (temporal) logics are the most used formalisms to specify the system and properties, respectively; which may be because they allow abstract specification, which is useful early in  development.

We found that model-checkers are the most often used verification approach, which complements the wide use of state-transition systems and temporal logics~\cite[Tables 3--4]{Luckcuck2019}. This may be because the model-checking approach is generally easy to explain to stakeholders with no experience of using formal methods. Notably, theorem provers were used a lot less often, we believe that this is due to the level of expert knowledge required to operate them correctly and efficiently.

\textbf{RQ3}, asked what are the limitations of the formalisms and approaches to verification that were identified in the answer to \textbf{RQ2} (see \cite[\S 7]{Luckcuck2019}). One obvious limitation appears to be a resistance to adopting formal methods in robotic systems development~\cite{Lopes2016}. The perception is that applying formal methods is a complicated additional step in the engineering process, which prolongs development while not adding to the value of the final product. A lack of appropriate tools also often impedes the application of formal methods. There are, however, notable examples of industrial uses of formal methods~\cite{Woodcock2009}.

There have been a variety of tools developed for the same formalism~\cite[Table 3]{Luckcuck2019} signaling a lack of interoperability between different formalisms and tools. Often, models or specifications of similar components are incompatible and locked into a single tool. Thus, a common framework for translating between, relating, or integrating different formalisms, would help smooth the conversion between formalisms/tools. This would also serve a growing need to capture the behaviour of complex systems using a heterogeneous set of formalisms and integrated formal methods. This is an open problem in formal methods for robotic systems\cite{Farrell2018}.

Another limitation faced in this domain is in formalising the \textit{last link}, the step between a formal model and program code. Ensuring that the program correctly implements the model requires a formalised translation. The lack of clarity about this limitation points to another: a lack of open sharing of models, code, and realistic case studies that are not tuned for a particular formalism. 

Field tests and simulations are both useful tools for robotic systems development\cite[\S 2]{Luckcuck2019}; but formal verification is crucial, especially at the early stages of development when field tests of the control software are infeasible (or dangerous). A focussed research effort on combining or integrating formal methods should improve their use in robotic systems development, because no single formalism is capable of adequately checking that all aspects of a robotic system behave as expected. Ensuring that these tools are usable by developers and providing similar features in an IDE would also improve their uptake by simplifying their use. Work in this area could lead to an Integrated \textit{Verification} Environment, allowing the use of different formalisms using same developer front-end, connecting them to their respective tools, and providing helpful IDE-like support.

\section{Conclusion}
The development of autonomous robotic systems is a novel, emerging, and quickly-changing field. Many of these systems are inherently safety- or mission-critical, so it is prudent that formal methods are used to ensure that they behave as intended. To advance research in this area, our survey provides a description of the formalisms and tools that are current being applied to autonomous robotic systems. To provide some guidance for choosing these languages and tools, we describe them and the case study tackled for the surveyed literature~\cite[Table 1]{Luckcuck2019}, however a detailed analysis of this is left as future work. The survey also highlights the shortcomings of these approaches and outlines exciting and necessary future directions for the entire formal methods community.

\bibliographystyle{splncs04}
\bibliography{bibliography} 

\end{document}